\documentclass[openacc]{rstransa}
\usepackage{aas_macros}

\titlehead{Opinion piece}

\begin{document}

\title{Cosmic mysteries and the hydrogen 21-cm line: bridging the gap with lunar observations}

\author{
A. Fialkov$^{1,2}$, T. Gessey-Jones$^{2, 3}$ and J. Dhandha$^{1,2}$ }

\address{$^{1}$Institute of Astronomy, University of Cambridge, Madingley Road, Cambridge, CB3 0HA, UK\\
$^{2}$Kavli Institute for Cosmology, Madingley Road, Cambridge, CB3 0HA, UK\\
$^{3}$Astrophysics Group, Cavendish Laboratory, J. J. Thomson Avenue, Cambridge, CB3 0HE, UK}
\subject{The 21-cm signal of hydrogen from cosmic dawn and dark ages}


\corres{Anastasia Fialkov\\
\email{afialkov@ast.cam.ac.uk}}

\begin{abstract}
The hydrogen 21-cm signal is predicted to be the richest probe of the young Universe including eras known as the cosmic Dark Ages, Cosmic Dawn when the first star and black hole formed, and the Epoch of Reionization. This signal holds the key to deciphering processes that take place at the early stages of cosmic history. In this opinion piece, we discuss the potential scientific merit of lunar observations of the 21-cm signal and their advantages over more affordable terrestrial efforts. The Moon is a prime location for radio cosmology which will enable precision observations of the low-frequency radio sky. The uniqueness of such observations is that they will provide an unparalleled opportunity to test cosmology and the nature of dark matter using the Dark Ages 21-cm signal. No less enticing is the opportunity to obtain a much clearer picture of Cosmic Dawn than what is achievable from the ground, which will allow us to determine the properties of the first stars and black holes.
\end{abstract}


\begin{fmtext}

\end{fmtext}


\maketitle
\section{The present-day landscape}

We are fortunate to live in an era of spectacular successes in observational cosmology. Large-scale imaging surveys like BOSS \cite{BOSS_BAO}, DES \cite{DES_2021}, DESI\cite{DESI_data} and the recently launched {\it Euclid} \cite{Euclid} are able to scan the nearby Universe in great detail and map out the positions of billions of galaxies. At the same time, intensity mapping experiments such as MeerKAT \cite{MeerKAT1, MeerKAT2}, CHIME \cite{CHIME_overview} and  Tianlai \cite{Tianlai} aim to provide complementary information by probing the large-scale distribution of neutral hydrogen in galaxies \cite{CHIME_HI, MeerKAT1, MeerKAT2}. Combined, all these surveys provide expansive maps of our cosmic neighborhood, covering the observable Universe all the way out to redshift $z \sim 3$, when the Universe was merely 2 Gyr old. In addition to these low redshift probes, observations of the Cosmic Microwave Background (CMB), e.g.\ with the {\it Planck} satellite, provide a comprehensive picture of the Universe when it was only 0.38 million years old ($z\sim 1000$) \cite{planck2018cosmo}.

The era between low-redshift observations and the much higher-redshift CMB last scattering surface is less well probed. Epochs such as Cosmic Dawn and the Dark Ages, forming the first few 100 million years of cosmic history, remain largely unobserved. These epochs host a large number of cosmological milestones and landmark astronomical events, such as the build-up of the first dark matter halos massive enough to hold gas, the birth of the first stars and black holes, and the onset of reionization of neutral gas by UV stellar photons which is often referred to as "the last phase transition" of the Universe \cite{Barkana2016}. The promise of exciting scientific discoveries sparks the enormous interest of the observational community in probing these epochs and motivates the launch and design of new telescopes.

However, the high-redshift Universe is notoriously difficult to observe. The high required sensitivity of galactic surveys \cite{MD,zackrisson2011spectral,Steinhardt2021}, presence of bright Galactic \cite{Galactic1998,Foregrounds2008} and extragalactic \cite{LOTSS,gleam2017} foregrounds in the radio sky (see \cite{chapman2019foregrounds} for a recent review), and systematics \cite[for example]{steinhardt2023templates,arrabal2023spectroscopic,hills2018concerns} prove to be a significant challenge. Despite this, the scientific community is actively pushing the observational frontier to earlier cosmic times. The recent launch and subsequent observations by the JWST have begun probing the bright galaxy population deep into the Epoch of Reionization at $z>10$, pushing the limits of its predecessor, the Hubble Space Telescope (HST). Large JWST fields such as CEERS \cite{finkelstein2023ceers}, GLASS and JADES have revealed hundreds of candidate galaxies at such early epochs, with the current record holder for the most distant spectroscopically confirmed object at $z\sim13.2$ being \mbox{JADES-GS-z13-0} \cite{curtis2023spectroscopic,robertson2023identification}. Furthermore, we are beginning to see the "monsters" inhabiting the early Universe: supermassive black holes  \cite{larson2023ceers,furtak2023supermassive}, high redshift quasars \cite{yue2023eiger,christensen2023metal} with a record-breaking X-ray luminous quasar UHZ1 at $z=10$ \cite{bogdan2023detection,goulding2023uncover}, and Active Galactic Nuclei (AGN) all the way out to the exceptionally luminous GN-z11 at $z=10.6$ \cite{tacchella2023jades,bunker2023jades}.

Despite these successes, we are merely probing the tip of the iceberg. The observations by the JWST leave out the most typical galaxies, which are dimmer than the threshold JWST sensitivity. Such galaxies are expected to be numerous and collectively may have had a strong influence on the state of the early Universe. The 21-cm signal of neutral hydrogen is expected to probe this population of galaxies by measuring their impact on the thermal and ionization histories. 

\section{The science-rich 21-cm signal}

The 21-cm signal of neutral hydrogen from the intergalactic medium (IGM) is predicted to be the most sensitive probe of the Universe at the Epoch of Reionization and Cosmic Dawn, and the sole probe of the Dark Ages. Once detected, this signal will provide a three-dimensional map of the Universe at the broad redshift range $z\sim 6-1000$ \cite{Furlanetto2006, Barkana2016, Mesinger2020} (note that contrary to the common assumption, although the signal at very high redshifts is weak, it is non-vanishing owing to the departure of Ly\,$\alpha$ color temperature from gas temperature \cite{FL2013,21cmfrom1000}), corresponding to redshifted radio signals at $\sim 1-200$ MHz frequencies.  

\begin{figure}[!ht]
\centering\includegraphics[width=5.4in]{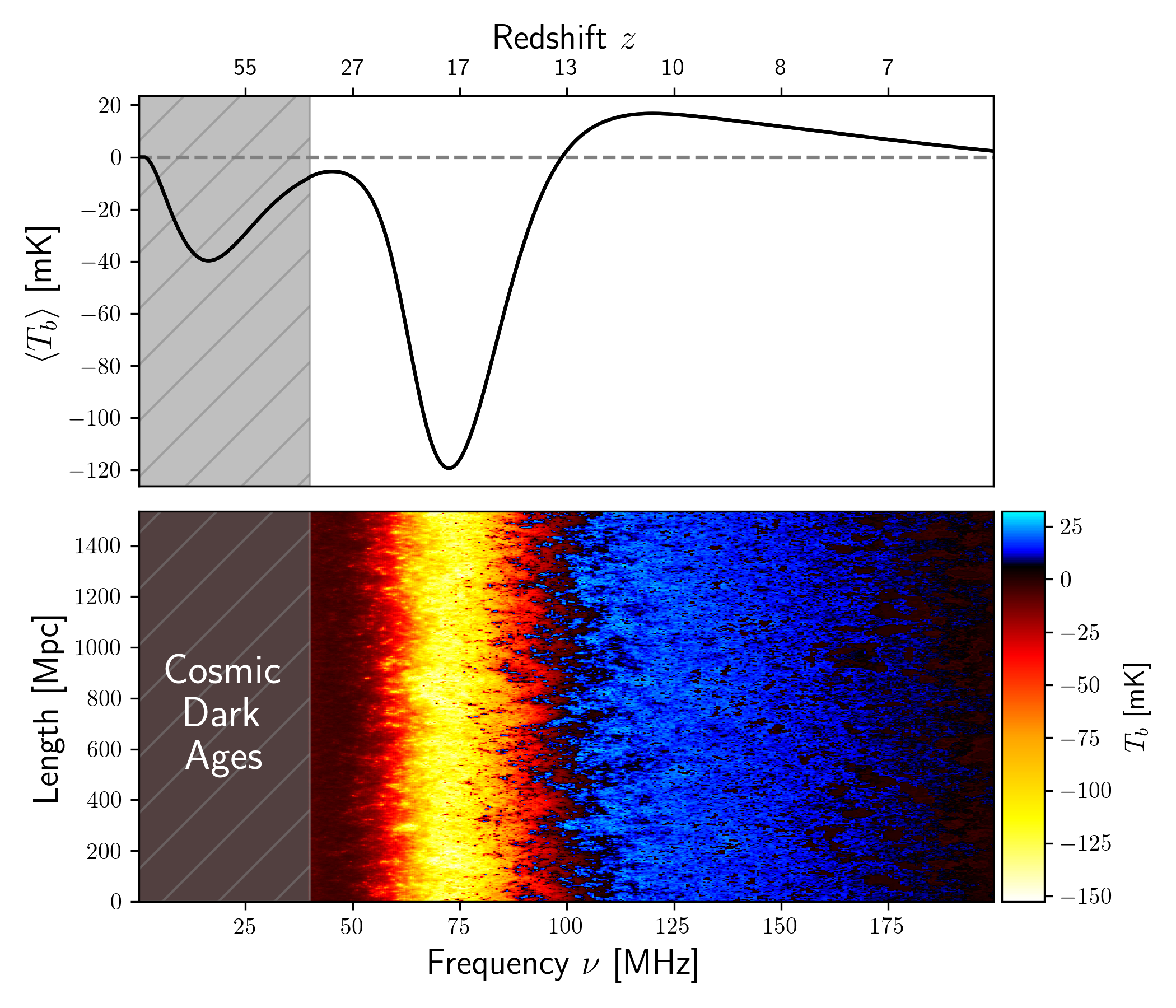}
\caption{The 21-cm signal across cosmic time. The demonstrated timeline covers (from left to right) the Dark Ages, Cosmic Dawn, and the Epoch of Reionization, with the process of reionization completed by $z\sim 5$. We show the sky-averaged (global) signal (top) and a lightcone map of spatial fluctuations (bottom) as a function of time (horizontal) and space (vertical). The post-Dark Ages signal is generated using the semi-numerical code {\sc 21cmSPACE} \cite{visbal2012signature,fialkov201321,fialkov2013complete,fialkov2014observable,fialkov2014rich,fialkov2015reconstructing,cohen201621,fialkov2017constraining,fialkov2018constraining,fialkov2019signature,reis2020high,reis2021subtlety,reis2022shot,magg2022effect,gessey2022impact,sikder2023strong,gessey2023signatures}, while the Dark Ages global signal is generated using the analytic code {\sc Recfast++} \cite[for model details]{acharya2022can}. The colorbar on the right shows the differential brightness temperature of the 21-cm line in mK. The simulations assume the standard $\Lambda$CDM cosmology with cosmological parameters from the Planck 2018 analysis~\cite{aghanim2020planck}. In this simulation, stars are assumed to form in halos with circular velocity above 4.2 km/s. The adopted astrophysical model assumes Population III star formation with fixed 0.2\% efficiency and log-flat initial mass function \cite{gessey2022impact}, an intermediate time-delay transition between Population III and Population II star-forming halos \cite{magg2022effect}, and Population II star formation with a simple power-law efficiency \cite{park2019inferring}. X-ray binaries are assumed to produce X-rays with efficiency $f_X=1$ and have soft SEDs with power-law exponent of $-1.5$ above 0.1 keV \cite{fialkov2014observable}, while galaxies produce radio emission with an efficiency of $f_\text{rad}=10$ \cite{reis2020high}. Cosmic ionization efficiency is assumed to be of $\zeta=15$ \cite{FurES}. The simulations also include various feedback processes such as Lyman-Werner feedback \cite{fialkov201321}, photo-heating \cite{cohen201621}, and baryon-dark matter relative motion \cite{visbal2012signature}.}
\label{fig_lightcone}
\end{figure}

This signal, demonstrated in Figure \ref{fig_lightcone}, is a rich probe of astrophysics and cosmology. The top panel shows the sky-averaged (or global) 21-cm signal which can be used to determine the timing of cosmic milestones (e.g. the onset of star formation, the moment when X-ray binaries re-heated the IGM to the temperature of the CMB, and the end of reionization). The bottom panel shows the lightcone, i.e. spatial and temporal structure of the signal. We see that the 21-cm signal is highly non-uniform at most stages of cosmic history with the fluctuation pattern changing in time as the Universe evolves and new processes dominate the signal.  The figure covers several key stages in the evolution of temperature and the ionization state of the IGM including (from left to right) the Dark Ages ($z\gtrsim 30$), Cosmic Dawn ($z\sim 10-30$), and the entirety of the Epoch of Reionization ($z\sim 10-6$, with the process of reionization completed by $z\sim 5$ in this specific simulation).


The Dark Ages 21-cm signal is largely determined by structure formation and dark matter physics. This signal probes fluctuations in the baryon density, peculiar velocity, and baryon temperature. Precision modeling of the  21-cm signal from the Dark Ages requires the inclusion of cosmological phenomena such as redshift evolution (light-cone), the Alcock-Paczynski effect, the relative velocity between dark matter and gas, collisions of hydrogen atoms with various species, the color temperature of the residual Ly\,$\alpha$ photons left over from Recombination, and the distribution of the mildly (at the level of $\sim 10^{-4}$) ionized gas \cite{lc2007, FL2013,  daVBC2014, 21cmfrom1000, mondal2023prospects}.

Collisional coupling between the hydrogen atoms and all existing particles throughout the cosmic dark ages guarantees that the 21-cm signal is driven by the kinetic temperature of the gas, resulting in a non-vanishing but faint radio signal visible in absorption against the CMB background \cite{Furlanetto2006}. Within the conventionally considered "standard" models of cosmology and astrophysics, the observable absorption feature has differential brightness $T_b \sim -40$ mK at $z\sim 75$ \cite[and Figure \ref{fig_lightcone} of this work]{mondal2023prospects}. This signal is typically much weaker than the Cosmic Dawn absorption trough which we discuss in the following paragraph  (centered at $z\sim 18$ for the model demonstrated in Figure \ref{fig_lightcone}), but is roughly of the same brightness as the reionization emission peak which occurs in models with strong enough heating (at $z\sim11$ for the model shown in Figure \ref{fig_lightcone}). Note that in scenarios with inefficient heating \cite{fialkov2014observable}, the neutral IGM might be colder than the CMB for the entirety of cosmic history (including the Epoch of Reionization) thus resulting in the 21-cm signal seen in absorption at any redshift (i.e. with no emission feature) and "cold reionization".  As the Universe expands and collisions become less efficient, the 21-cm signal fades away becoming practically undetectable (around  ${\mbox z\sim 30}$ in Figure \ref{fig_lightcone}).

The next milestone in the history of the Universe is the formation of the first sources of light which ushers the Universe into the Cosmic Dawn era ($z\sim10-30$).
As the first stars emerge at $z\sim30$ \cite{klessen2023first}, they produce Ly $\alpha$ photons that couple the 21-cm spin temperature to the kinetic temperature of the gas \cite{Wouthuysen1952, Field1958, fialkov201321,fialkov2013complete, gessey2022impact}. Owing to the different adiabatic cooling rates, the gas temperature is colder than the CMB, resulting in an observable absorption signal. As the Universe expands and adiabatically cools down further, the absorption deepens -- a process that continues until the first population of heating sources emerges. The onset of cosmic heating defines the so-called "absorption trough", the global minimum of the sky-averaged 21-cm signal, clearly seen in the top panel of Figure \ref{fig_lightcone} at ${\mbox z\sim18}$. One of the most widely considered types of heating sources are the first X-ray binaries \cite{fragos, sartorio2023population}. These astrophysical objects are natural end-points of stellar evolution \cite{MadauRees2001, HaimanLoeb2001}; as the first stars die, some end up as compact objects in binary systems (e.g. the first astrophysical black holes). These systems produce X-rays in the process of accretion \cite{sartorio2023population} or decretion \cite{liu2023population} of gas. The X-ray background contributes to reheating of the IGM in a non-uniform manner (see Figure \ref{fig_lightcone}, \cite{PF2007, fialkov2014observable,fialkov2017constraining}) and the contrast between the gas temperature and the CMB decreases. The gas can also be heated by other astrophysical sources, e.g. via cosmic rays \cite{SS2015, gessey2023signatures} or Ly $\alpha$ scattering \cite{Chen2004, Chuzhoy2007, reis2021subtlety}. As we mentioned above, depending on the efficiency of the first heating sources, the neutral gas temperature might either rise above that of the background radiation resulting in an emission 21-cm signal (as shown in Figure \ref{fig_lightcone} at the low-redshift end) or remain colder than that of the CMB resulting in an absorption 21-cm signal until the end of reionization \cite{fialkov2014observable}. Finally,  the signal vanishes as the neutral hydrogen in the IGM is ionized by galaxies and quasars.

The Cosmic Dawn global 21-cm trough ($z\sim 18$ in Figure \ref{fig_lightcone}) might be a few hundred mK deep \cite[with the exact location and depth being model-dependent]{Cohen2017, fialkov2019signature, reis2021subtlety} and the spatial structure of the signal is predicted to have a rich fluctuation pattern that could inform us on some of the earliest astrophysical processes \cite{BL2005, PF2007,  fialkov2014observable, gessey2023signatures}. This deep absorption is the target of many ground-based missions such as the radiometers EDGES~\cite{bowman2018absorption}, MIST~\cite{monsalve2023mapper}, REACH~\cite{acedo2022reach}, and SARAS~\cite{singh2022detection}, and the interferometers which target fluctuations in the 21-cm signal such as  HERA~\cite{abdurashidova2022first}, LOFAR~\cite{Gehlot2019, mertens2020improved}, NenuFAR~\cite{mertens2021exploring, Nenufar2023}, MWA~\cite{ewall2016first}, LWA \cite{LWA1}, LEDA \cite{LEDA} as well as the future SKA~\cite{dewdney2009square,koopmans2015cosmic}. 

In addition to the {\it commonly considered} astrophysical and cosmological processes described above, the signal will depend on other processes if they affect the growth of structure, star and black hole formation, or heating and ionization of the Universe. For instance, dark matter cooling~\cite{Barkana2018, Munoz2018, fialkov2018constraining, Kovetz2018, Liu2019} or excess radio background above the CMB level \cite{Feng2018, EW2018, fialkov2019signature, EW2020, reis2020high, sikder2023strong}  will affect the structure, magnitude and features of the signal.

\section{Science with ground-based 21-cm observations}

The science-rich 21-cm signal outlined above is hard to measure owing to its intrinsic faintness, the brightness of overlaying foreground signals, and the uncertainty in instrumental systematics.

Terrestrial observations of the Dark Ages are made particularly difficult by the ionosphere which corrupts low radio frequencies. Due to the electromagnetic properties of the ionosphere,  signals at frequencies below $\mathcal{O}(10)$ MHz (plasma frequency of the F-layer peak) are reflected into space and cannot be observed from the ground, while radio waves below $\sim 300$ MHz are refracted and partially absorbed  \cite{Lusee, Iono, VLK2014}. As a result, the cosmic Dark Ages, which are encoded in the lowest radio frequencies owing to the expansion of the Universe, can only be measured from above the ionosphere providing the most compelling science case for lunar and space missions.

\begin{figure}[!ht]
\centering
\includegraphics[width=3in]{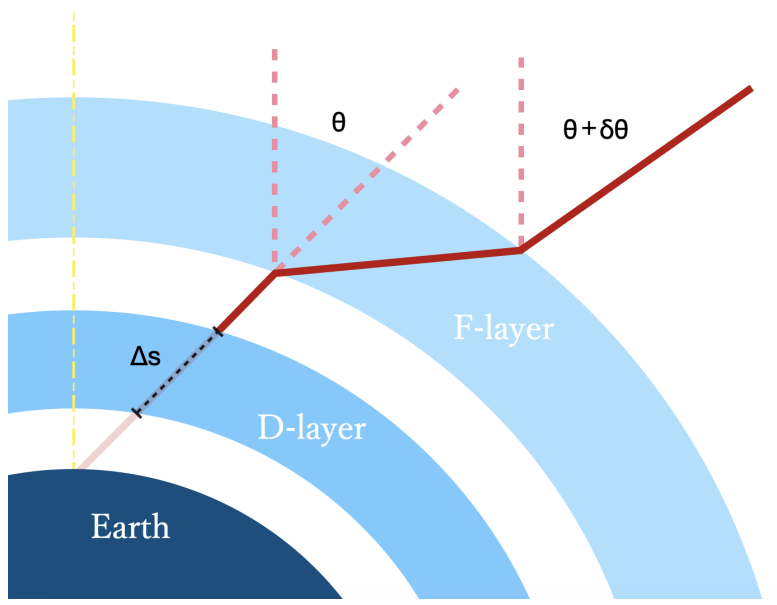}\includegraphics[width=2.2in]{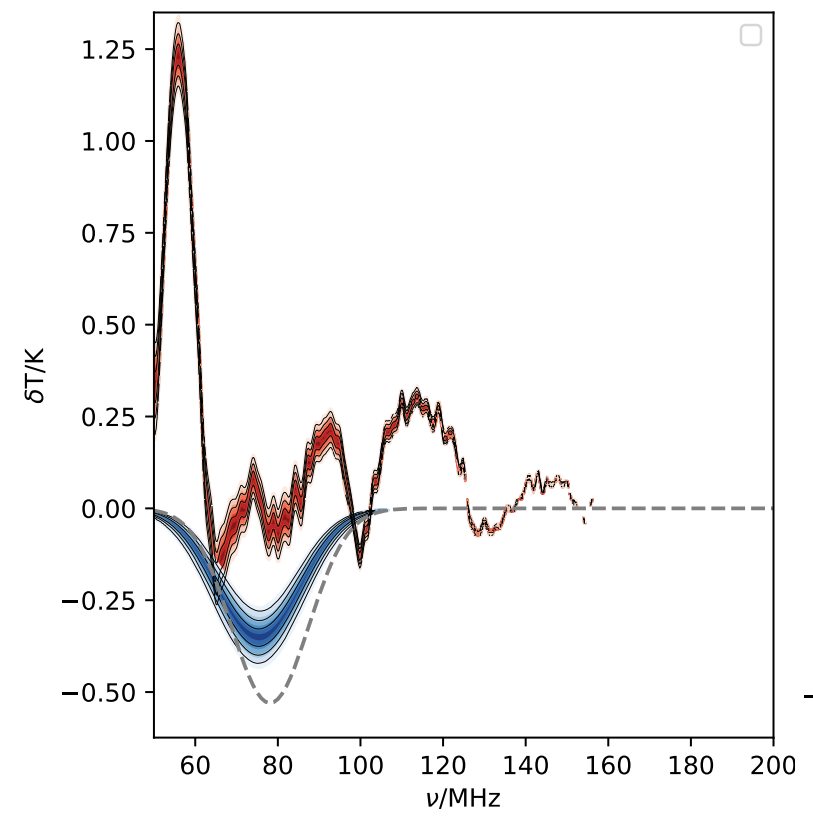}
\caption{{\bf Left:} Refraction and absorption of homogeneous ionospheric layers, not-to-scale, adapted from  \cite{shen2021quantifying}. The illustration shows two layers (F-layer and D-layer) of the ionosphere that have the most important impact on radio wave propagation: the F-layer is the highest region of the ionosphere and has the highest density of free electrons thus providing the dominant contribution to refraction. Although its average degree of ionization does not vary significantly through the night, the ion distribution might vary. The D-layer dominates absorption. {\bf Right:} Measuring the global 21-cm signal in the presence of time-varying ionosphere. The signal is extracted using a Bayesian pipeline of the REACH experiment \cite{REACHpipeline2021}. The data includes an injected/{\it true} cosmic signal (grey dashed), residuals with the fitted foreground removed are shown in red, and the best-fit reconstructed global 21-cm signal posterior is shown in blue. Plot adapted from Shen et al. 2022 \cite{shen2022bayesian}. Time-varying ionosphere is implemented using the real data collected from Lowell GIRO Data Center at station Louisvale, South Africa. Shen et al. 2022 conclude that more than 5\% error in the ionospheric (time-dependent) model will impede the global 21-cm measurement.}
\label{fig_moon}
\end{figure}

Although the remaining parts of cosmic history, including the signals from Cosmic Dawn and the Epoch of Reionization, are accessible from the ground, the ionosphere acts as a lens creating chromatic distortions of the incoming low-frequency radio waves \cite[and right panel of Figure \ref{fig_moon}]{VLK2014, shen2021quantifying}. Shen et al. \cite{shen2022bayesian} showed that more than  5\% error in a time-dependent ionospheric model will corrupt the global 21-cm measurement (left panel of Figure \ref{fig_moon}). The ionosphere, naturally, also creates a problem for interferometric observations of fluctuations in the 21-cm signal. Ionospheric propagation delays are a major contributor to phase errors at low radio frequencies and can pose a significant challenge even for the Epoch of Reionization experiments \cite{Iono2}. Although currently, the impact of the ionosphere is often left untreated, the ionospheric effects can be (at least partially) removed, e.g. LOFAR is using direction-dependent calibration \cite{Iono3}. 

In addition to the ionospheric distortions, human-made radio frequency interference (RFI) contaminates the signal making the ground-based observations harder to interpret. These issues lead us to think that lunar observations of Cosmic Dawn and the Epoch of Reionization  (especially from the lunar dark side to avoid RFI) would provide a much clearer view of the epochs and allow us to robustly extract some of the most interesting details of primordial star and black hole formation as well as shedding light on the nature of dark matter and structure formation at early times.

Despite the difficulties, many of the existing ground-based low-frequency radio telescopes provide competitive upper limits that are, in some cases, strong enough to rule out most extreme theoretical models.

A fully Bayesian analysis conducted by Bevins et al. 2023 \cite{bevins2023Joint} showed that, at the time of writing, 
HERA provides the tightest constraints on the 21-cm power spectrum from the Epoch of Reionization \cite{adams2023improved}, followed closely by LOFAR \cite{ghara2020constraining} and MWA \cite{trott2020deep}.  
The latest publicly available HERA limits (at 95\% confidence) are $\Delta^2 = 457$ mK$^2$ at $k = 0.34$ h Mpc$^{-1}$ and $z = 7.9$, and $\Delta^2 = 3496$ mK$^2$ at $k = 0.36$ h Mpc$^{-1}$ and $z = 10.4$, derived using 94
nights of observing with HERA Phase I.  In addition to the constraints obtained using the interferometric data, residuals of the global signal experiments SARAS 2 and EDGES High-Band appear to be low enough to rule out some of the standard astrophysical scenarios at the reionization redshifts \cite{SARAS2, Monsalve2017, Monsalve2018a, Monsalve2019a}. 

Observations of Cosmic Dawn are more controversial. The EDGES collaboration reported a tentative detection of a deep absorption trough at $z\sim 17$ with the EDGES Low-Band antenna \cite{bowman2018absorption}. This detection has not been confirmed and is in tension with SARAS 3 measurements at  {\mbox{$z\sim 15-25$}}  \cite{singh2022detection}. Exploration of Cosmic Dawn is also being conducted with interferometers including the 'AARTFAAC Cosmic Explorer' (ACE) program of LOFAR \cite{gehlot2020aartfaac}, NenuFAR \cite{Nenufar2023}, MWA \cite{MWA_highz}, LWA \cite{LWA1} and LEDA \cite{LEDA}. However, the published  Cosmic Dawn power spectra limits are very weak and do not constrain any astrophysical scenarios.

The constraining data (e.g. from HERA at $z = 8$ and $10$ \cite{adams2023improved} and SARAS 3 at $z \sim 15-25$  \cite{singh2022detection}) are being used to restrict standard and exotic astrophysical scenarios, including models with enhanced 21-cm signals boosted by the extra radio background present in addition to the CMB \cite{fialkov2019signature, reis2020high, sikder2023strong}.  Such models, originally designed to explain the anomalous EDGES Low-Band detection, provide an interesting theoretical test case. In a Bayesian analysis, limits on the 21-cm power spectrum at $z\sim 8$ and $10$ from HERA and global signal constraints at $z \sim 15-25$ from SARAS 3 were shown to limit the astrophysical parameter space of these models \cite{bevins2023Joint}. Bevins et al. (2023) \cite{bevins2023Joint} showed that in synergy, the two experiments leave only 
$64.9^{+0.3}_{-0.1}$\% of the explored prior space to be consistent with the joint data set. The strongest joint constraints are in the space of the radio and X-ray luminosities of the first galaxies. The joint analysis disfavors at 68\% confidence a combination of galaxies with X-ray emission that is  $\lesssim 33$ and radio emission that is $\gtrsim 32$ times as efficient as present-day galaxies. In addition, weak trends in constraints of star formation efficiency and minimum halo mass for star formation are seen.

The synergetic constraints by HERA and SARAS 3 can be further supplemented by the unresolved X-ray background measurements from the {\it Chandra} X-ray satellite \cite{Hickox2006, HarrisonXrays} and the radio background detected by ARCADE2 \cite{Fixsen2011}  and LWA1 \cite{Dowell2018}. In their work Pochinda et al. \cite{Pochinda2023} considered a model that differentiates between the primordial stars (Population III, see more discussion in Section 4 \ref{secCD} below) formed out of chemically pristine gas (at the Big Bang nucleosynthesis level) and second-generation stars (Population II) formed out of chemically-enriched gas. This study indicates that SARAS 3 data is (weakly) sensitive to the properties of Population III star-forming regions, while the other experiments mostly constrain the properties of X-ray and radio sources. Although very weak and model-dependent, these limits are one of the first to test the properties of primordial star-forming regions. This analysis illustrates that even the existing data, despite being plagued by systematic effects, ionospheric distortions and foreground uncertainties, can be used to advance our understanding of astrophysics at Cosmic Dawn.

\section{Science from the Moon}

Observations from the lunar surface or space will provide the only way to probe the state of neutral hydrogen during the Dark Ages as this radio signal is inaccessible from the ground. Moreover, these observations are expected to supersede terrestrial observations of Cosmic Dawn bypassing the issues of ionospheric distortions and radio frequency interference (if performed from the lunar dark side). Owing to the cleaner radio environment, measurements of the Cosmic Dawn 21-cm signal from the Moon might allow us to test some of the most intriguing properties of first stars and black holes. 

\subsection{Crisp observations of Cosmic Dawn}
\label{secCD}

Precise 21-cm signal measurements of Cosmic Dawn are probably our best chance to probe the first generation of stars (also called Population III or Pop III stars) and the successive population of first X-ray binaries (XRBs). Little is known for certain about these objects~\cite{klessen2023first, sartorio2023population}, but it is widely believed that the first metal-free stars form in small numbers in dark-matter mini-halos from the hydrogen and helium gas produced in Big Bang nucleosynthesis. Despite making up only a tiny fraction of the stars that will ever form, these stars should have had a profound impact on the history of the Universe, producing the first metals, and starting the reionization of the IGM. Properties of the first population of XRBs, which are responsible for the onset of the IGM heating, are tightly linked to the properties of the stars themselves (such as the stellar initial mass function (IMF) \cite{sartorio2023population}).  

The sensitivity of the Cosmic Dawn 21-cm signal to the first stars and XRBs, via the Ly\,$\alpha$ photon emission of the former and X-rays generated by the latter, allows it to probe these first sources of light.
At the most basic level, detecting global features associated with Cosmic Dawn in the 21-cm signal, like the rapid drop in the 21-cm global signal and the subsequent rise as a result of heating (demonstrated in Figure~\ref{fig_lightcone}), will reveal the timing and efficiency of the formation of the first stars and XRBs. Further high signal-to-noise measurements of the 21-cm signal from Cosmic Dawn should provide additional details and insights into the properties of the first sources.

\begin{figure}[!ht]
\centering\includegraphics[width=5.2in]{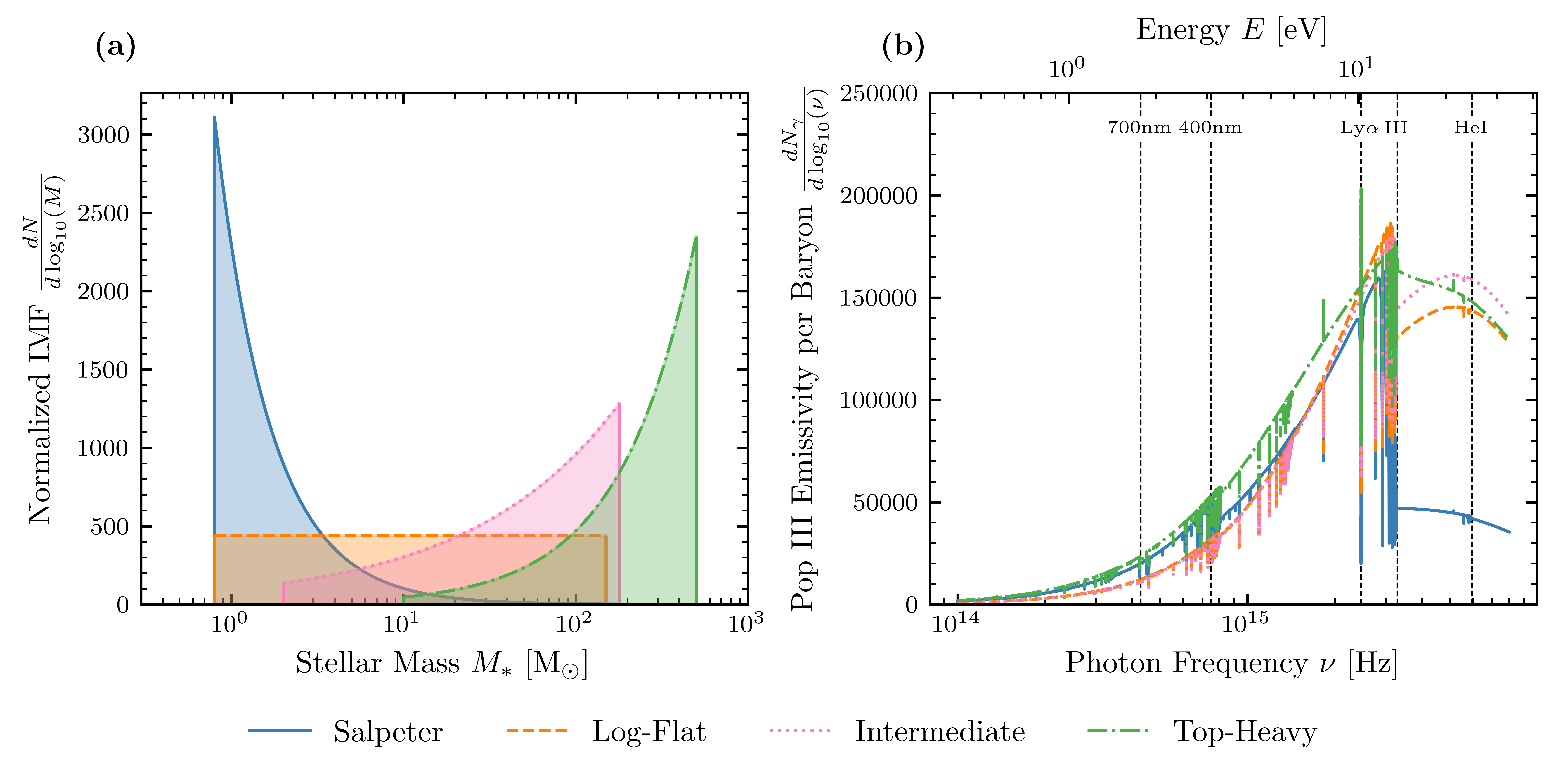}
\caption{The emissivity of the first stars varies with their mass distribution. Panel (a, left) depicts four truncated power-law initial mass functions (IMFs) that may describe the mass distribution of the first stars. Panel (b, right) shows the corresponding prediction for the population-averaged emissivity per baryon of the first stars given these IMFs. The {\mbox{Lyman-$\alpha$}} line (Ly\,$\alpha$), hydrogen ionizing energy (HI), and helium ionizing energy (HeI) are highlighted as the emissivity above these lines has particular importance in determining the evolution of the 21-cm signal. Moderate differences are seen in emission between Ly\,$\alpha$ and HI between IMFs and significant variations in ionizing emissivity. The Lyman band emissivity variations produce potentially observable differences in the 21-cm signal \cite{gessey2022impact}, suggesting the possibility of probing the mass distribution of the first stars using the 21-cm signal. }
\label{fig_sed}
\end{figure}

For example, emissivity of  a Pop III star in the Lyman band depends on the mass of the star \cite{Schaerer2002, gessey2022impact}. As a result, the combined signature of the first stellar population in the 21-cm signal depends on the distribution of stellar masses,  the so-called initial mass function (IMF, see Figure~\ref{fig_sed}). Different IMFs result in small, but potentially measurable, variations in the predicted Cosmic Dawn 21-cm signal~\cite{gessey2022impact}. 
If first star formation was efficient, these signatures may be just measurable by the SKA~\cite{koopmans2015cosmic}, though at low signal-to-noise ratios. However, if first star formation is inefficient or occurs earlier than anticipated (e.g. in the case of rare overdense regions) these differences will require precise low-frequency 21-cm signal measurements that are only feasible from the lunar dark side.

\subsection{The unseen Dark Ages}
The ability to probe the Dark Ages provides a unique science case for space and lunar observations of the 21-cm signal. Uncontaminated by astrophysics, it provides a new probe of fundamental physics over the unprecedentedly large range of scales and unseen cosmic time. By performing a mode counting exercise, Cole and Silk (2021) \cite{CS2021} found that lunar observations of the three-dimensional (nearly linear) 21-cm power spectrum from $z=50$ will probe $\sim 10^{12}$ modes, which is considerably larger than the amount of information contained in the CMB ($\sim 10^6$ modes) and large scale structure ($10^8$ modes at $z=1$).

The Dark Ages are marked by the first-ever infall of gas into newly assembled deep dark matter potential wells and, thus, provide an unprecedented opportunity to study the birth-places of first stars and the onset of structure formation \cite{BL2005, clumpy2010}. Low-frequency radio experiments with arcminute angular resolution will be able to probe non-Gaussianity produced by nonlinear collapse e.g. using the 21-cm bispectrum \cite{Pillepich2007}.  Beyond these exciting prospects is the highly compelling case of early cosmology \cite{mondal2023prospects}.
The large number of linear modes probed by the Dark Ages 21-cm signal will provide an unparalleled test of primordial non-Gaussianity of the initial density field \cite{Cooray2006, Pillepich2007, Floss2022, Orlando2023}. The signal from $z\sim 30-100$ will test the inflationary paradigm on small scales (down to $\sim 0.1$ Mpc) inaccessible to the CMB experiments, allowing us to probe theories with primordial non-Gaussianity of $f_{\rm NL}\gtrsim 10^{-2}$ \cite{Pillepich2007, Munoz2015, Floss2022}. This constraint improves ({\mbox{$f_{\rm NL}^{\rm loc}\sim 6\times 10^{-3}$}}) when cross-correlations between 21-cm fluctuations and the CMB T- and E-mode anisotropies are considered \cite{Orlando2023}.


If deviations of the Dark Ages 21-cm signal from the predictions of the standard $\Lambda$CDM cosmology are observed, this could be a signature of dark matter physics \cite{Tashiro2014, DMannihilation, Slatyer2018, Short2020, dasignaldm2023, da_AXIONS2023}, or other exotic processes \cite{Mack2008, strings2021, PBH2022, running2011, JitesPaper}. In any theory in which new phenomena contribute to structure formation, heating or ionization at early times, these signatures will be directly imprinted in the 21-cm signal. Some examples include ultra-light axions which would affect the 21-cm signal by changing the matter power spectrum and thus affecting the collapse of early structures \cite{da_AXIONS2023}, dark matter annihilation or decay which could change temperature and ionization state of the gas \cite{DMannihilation, dasignaldm2023}, primordial black holes which impact the signal through Hawking radiation (evaporation) or emitting radiation in the process of accretion \cite{Mack2008, PBH2022, JitesPaper},  and superconducting cosmic strings (e.g. \cite{strings2021}).

\subsection{Caveats}

Undoubtedly, the scientific merits of the low-frequency radio observations from the lunar farside are great, with exciting prospects to probe fundamental physics, cosmology, and high-redshift astrophysics. 
The advantages of dark-side lunar observations are clear: the lack of ionosphere and RFI, as well as environmental stability during the two-week lunar night which permits long uninterrupted integration times\cite{Burns2021}.
However, such observations are technically challenging (see the White Paper by Koopmans et al. (2021) \cite{Koopmans2021} for more details). 

Low-frequency radio astronomy is plagued by the presence of bright foregrounds \cite{Foregrounds2008} which are several orders of magnitude stronger than the intrinsically weak 21-cm signal. This problem is a challenge for 21-cm observations from the ground and the Moon alike. The foregrounds are stronger at lower frequencies and, therefore, will be a more serious obstacle for the robust identification of the Dark Ages 21-cm signal compared to the Cosmic Dawn or Reionization eras. A viable solution is to marginalize over the foreground parameters when inferring cosmological properties \cite{REACHpipeline2021}.

Operating from the Moon also involves unfamiliar technical challenges \cite{Burns2021, Koopmans2021}. For example, reflections from the lunar subsurface are not well understood \cite{SEP2018} and could corrupt the observation if not modeled adequately. Physical properties of the lunar regolith such as density and porosity \cite{gamsky2010physical} could impact mission operation. Other environmental challenges include the large temperature gradients ($\sim 100^{\circ}$C during the day and $-170^{\circ}$C at night)  which can destabilize instrumentation, the charged lunar dust environment \cite{grun2011lunar}, and micrometeoroid flux which can affect the longevity of experiments. For interferometry, to achieve the required high spatial resolution and sensitivity it is estimated that an array of $10^5$  individual antennas distributed over 100 km$^2$ is needed  \cite{Wolt2013, Koopmans2021, mondal2023prospects} with integration times of up to 10,000 hours required for precision cosmology \cite{mondal2023prospects}. To host such large experiments, very few suitable shadowed craters exist on the lunar farside \cite{le2023lunar} adding to the need for urgent international policy in protecting these environments for astronomical research.

\vspace*{-5pt}

\section{Conclusion}

In this opinion piece, we reviewed the physics and the observational status of 21-cm cosmology. We provided a short discussion of the scientific merit of lunar observations at low radio frequencies arguing that the Moon, indeed, is expected to provide a unique environment for 21-cm cosmology. A plethora of exciting scientific questions can only be answered by doing 21-cm science from space/lunar surface.  In particular, observations of the 21-cm signal from the cosmic Dark Ages are only possible from either space or the Moon. These unique measurements will open up a new window to study cosmology and structure formation in the unexplored regime when the first bound dark matter objects (e.g. halos) were forming. If measured, deviations from the predictions assuming the standard $\Lambda$CDM cosmology could point to the presence of "exotic" processes such as non-cold dark matter. 
Additionally, free from ionospheric distortions and human-made interference, observations from the Moon/space are also expected to provide a clearer view of the epoch of the first star and XRB formation than what is possible from the ground, enabling precision science at Cosmic Dawn. Specifically, robustly determining the typical masses (and hopefully the full mass distribution) of the first generation of stars is one of the most exciting scientific questions that can be answered with 21-cm cosmology from the Moon/space.

\ack{We thank the two anonymous referees for their constructive comments which helped improve the paper. AF is grateful to the Royal Society for its continuous and generous support in the form of a URF. TGJ acknowledges the support of the Science and Technology Facilities Council (STFC) through grant number ST/V506606/1. JD acknowledges support from the Boustany Foundation and Cambridge Commonwealth Trust in the form of an Isaac Newton Studentship. AF thanks C. J. O'Connell for the careful proofreading of this manuscript.}


\bibliographystyle{RS}
\bibliography{library_trunc}

\end{document}